\documentclass{emulateapj}
\usepackage{natbib}
\usepackage{graphicx}
\usepackage{amsmath}
\usepackage{amssymb}

\newcommand{\mhost}{\mbox{$M_{\rm host}$}}
\newcommand{\msat}{\mbox{$M_{\rm sat}$}}
\newcommand{\rram}{\mbox{$r_{\rm ram}$}}
\newcommand{\rsoi}{\mbox{$r_{\rm soi}$}}
\newcommand{\msun}{\mbox{$M_{\odot}$}}
\newcommand{\mgal}{\mbox{$m_{\rm gal}$}}
\newcommand{\dd}{\mbox{${\rm d}$}}
\newcommand{\tdm}{\mbox{$\left<t\right>_m-dm_{\rm loss}'$}}
\newcommand{\tm}{\mbox{$\left<t\right>_m$}}
\newcommand{\dmloss}{\mbox{$dm_{\rm loss}'$}}
\newcommand{\fpass}{\mbox{$f_{\rm pass}$}}
\newcommand{\fearly}{\mbox{$f_{\rm early}$}}

\begin{document}

\shorttitle{RSF in satellites}
\shortauthors{Taysun Kimm et al.}

\title{The Impact of Gas Stripping and Stellar Mass Loss on Satellite Galaxy Evolution}

\author{Taysun Kimm$^{1,2}$, Sukyoung K. Yi$^{1}$, \& Sadegh Khochfar$^{1,3}$}
\affil{$^1$ Department of Astronomy, Yonsei University, Seoul 120-749, Korea}
\affil{$^2$ Currently at Astrophysics, University of Oxford, OX1 3RH}
\affil{$^3$ Max-Planck-Institut f\"{u}r Extraterrestial Physics, Giessenbachsrasse 1, 85748 Garching, Germany}

\begin{abstract}
Current semi-analytic models (SAMs) of galaxy formation over-predict the fraction of
passive small late-type satellite galaxies in dense environments by a factor of two to three.
We hypothesize that this is due to inaccurate prescriptions on cold gas evolution.
In the hope of solving this problem we apply detailed prescriptions
on the evolution of diffuse hot gases in satellites and on stellar mass loss,
both of which are critical to model cold gas evolution. 
We replace the conventional shock-heating motivated instant stripping with a realistic 
gradual prescription based on ram pressure and tidal stripping. 
We also carefully consider stellar mass loss in our model. When both 
mechanisms are included, the fraction of passive late types matches the data much more closely. 
The satellite over-quenching problem is still present in small galaxies in massive haloes, however.
In terms of the detectable residual star formation rates, gradual diffuse gas
stripping appears to be much more important than stellar mass loss in our model.
The implications of these results and other possibilities, such as redshift-dependent 
merging geometry and tidal disruption, are also discussed. 
\end{abstract}

\section{Introduction}

In the LCDM model, dark matter structures grow via merging and 
accretion. Within these dark matter halos, baryons initially
agglomerate onto the centre of the potential well within the free-fall 
time of the halo and form stars \citep{rees77,binney77,silk77}. 
When the halo becomes large enough, the accreted gas can be 
gravitationally shock-heated and turn into diffuse hot halo gas. 
In the meantime, the constant attraction of gravity causes haloes 
and their galaxies to interact and merge, forming clusters and groups 
of galaxies. While orbiting within a host halo, satellite galaxies are 
likely to interact with other galaxies and the hot ambient gas \citep[e.g.][]{chung07,yagi10}. 
The evolution of the cold and hot gas contents is heavily influenced 
by the details of these interactions, and the exact mechanism is not well understood.

It is known that galaxies in denser environments are optically redder 
and more quiescent than field galaxies 
\citep[to cite a few][]{gisler78,larson80,dressler80}.
This strongly suggests that the cold gas supply (the source of 
star formation) is controlled by mechanisms closely associated with 
the environment. 
\citet{gunn72} demonstrated that the cold gas of a galaxy can be stripped 
off due to the ram pressure exerted on it by the intracluster medium 
as a galaxy moves within a cluster 
potential \citep[see also][]{abadi99,quilis00,chung07,tonnesen09,yagi10}.
Tidal stripping is another important process that indirectly 
affects the evolution of cold gas. Satellites trapped by a larger 
cluster halo are tidally stripped off their dark haloes and hot gas reservoir.

Throughout their orbital motions, satellites lose their interstellar/diffuse halo gas 
through shock heating, ram pressure and tidal stripping of their dark halo.
Consequently, star formation in satellites may decreases with time. 
Semi-analytic models (SAMs) of galaxy formation assume that the shock 
heating is very efficient, and it is generally assumed to {\it instantly} 
remove diffuse halo gas from satellite haloes  \citep[e.g.][]{kauffmann99}.
Since the hot gas reservoir is removed instantly, gas cooling stops,
cold gas is quickly depleted through star formation and SN feedback,
and eventually star formation is quenched.
Such models predict that the bulk of satellite galaxies in large
haloes should be virtually passive, with only about 20\% predicted 
to be active \citep[e.g.][hereafter K09]{kimm09}. 

However, this prediction is not supported by observation.
Studies based on the {\em specific star formation rates} measured from 
{\it GALEX} and {\it SDSS} observations \citep{salim07} have found that
a much larger fraction of satellite galaxies are 
active \citep{kimm09}. 
This is sometimes referred to as the {\it satellite over-quenching problem} 
\citep[e.g.][]{weinmann06,baldry06,vdb08,gilbank08,fontanot09,kimm09}.

The assumption of instantaneous shock heating of satellite gases is thought to be 
the most likely cause of this mismatch between model and data.
Observationally, there is evidence for hot gas around satellite galaxies, 
as shown by {\it Chandra} observations of early-type galaxies in nearby 
clusters \citep{sun07,jeltema07}. These observations are bolstered 
by recent hydrodynamical simulations suggesting that a non-negligible 
fraction (up to $\sim$ 40\%) of diffuse hot gas remains associated with  
satellite galaxies for several Gyrs after merging with the host 
halo \citep{bekki02,kawata08,mccarthy08}. Various SAMs have included 
the effect of ram pressure stripping \citep{khochfar08,kang08,font08,guo10}.
In a recent attempt \citet{font08} showed that inclusion of ram pressure 
stripping can help improve the bimodal colour distribution of satellite galaxies 
and the dependence of satellite colours on galaxy environment \citep[see also][]{kang08}. 
However, they did not take into account the decay of the satellite 
orbits due to dynamical friction, and hence their calculations may slightly 
overestimate the amount of hot gas retained in the satellite systems. 
Further, as we will show, the instantaneous recycling approximation 
for stellar mass loss used in the model may change the recent star 
formation history of galaxies. Alternatively, \citet{weinmann10} propose 
a simple {\em gradual} recipe for the stripping of gas and dark matter 
that reproduces the specific star formation rates as a function of clustocentric radius.

Another factor that can contribute to the interstellar medium is 
 stellar mass loss \citep[e.g.][]{bregman09}. 
Stellar masses vary widely and their lifetimes depend strongly
on their mass. Thus, the stellar mass loss of a population is not instantaneous
but changes gradually with time (though the overall change can be dramatic). 
Yet, most SAMs  (\citealt{hatton03} is a notable exception) 
have used a simple approach in which roughly 30\% - 40\%
of newly formed stars {\em instantaneously} evolve off and are 
recycled to cold gas \citep[e.g.][]{somerville08}.
This may be a reasonable approximation for long-timescale 
star formation, but it may not be adequate for  
investigating short-timescale phenomena, such as the recent star 
formation history of satellite galaxies.

In this study, we aim to investigate the effects of the two physical
processes mentioned above in the context of SAMs. 
We first consider the orbital motion of a satellite under the influence 
of dynamical friction and the associated tidal stripping, both of which 
have been neglected in the previous studies based on EPS \citep[e.g][]{lacey93} 
formalism\footnote{SAMs based on N-body merger trees can follow the tidal 
stripping of dark matter subhaloes as long as they are robustly identified. 
Beyond that point, satellites are tracked based on analytic formalism.}. 
We also include a prescription for ram pressure stripping (\S 2.1). 
Next, in \S 2.2, we present further improvements on the earlier implementations of stellar mass loss.
We finally describe the impact of these considerations on satellite
galaxy evolution in \S 3 and discuss the implications of this in \S4.

\section{Semi-Analytic Model}
We adopt a semi-analytic approach of galaxy formation to investigate 
the effect of different physical processes on the recent star formation 
history of satellite galaxies. The fiducial SAM we are using 
is described in detail in \citet{khochfar05,khochfar06}. 
In what follows we briefly lay out the basic ingredients of the 
SAM and explain the important modifications to the fiducial model.

We generate dark matter halo merger trees for a cosmological 
volume of $10^6 {\rm Mpc}^3$ using a Monte-Carlo method developed by 
\citet{somerville99}. Within these halos gas cools and collapses at the 
centre of each halo, forming a rotationally-supported disc galaxy. 
Star formation in this disc is modeled assuming a Schmidt-Kennicutt 
law \citep{kennicutt98}. Some massive stars quickly die as supernovae 
and release their energy, which heats up the cold gas in the disk.
Note that our fiducial model does not include recycling of mass loss
by stellar evolution.
When two haloes merge,  the most massive galaxy 
in the more massive halo is set to be the ``central'' while the other galaxies become 
"satellite galaxies". These satellites begin to fall toward the central galaxy 
due to dynamical friction and eventually merge with the central galaxy. 
If the mass ratio ($=\mhost/\msat$) between two galaxies is less than 3.5, 
we assume that the existing stars in both galaxies 
and the new stars formed by merger-driven starbursts build up
the spheroidal component of the merger remnant. 
It should be noted that galaxy mergers are the only way 
to form a bulge in our model \citep[c.f][]{athanassoula08}.
We use a mass ratio-dependent burst efficiency to determine the number 
of stars formed during the merger, drawn from hydrodynamic simulations \citep{cox08}.
As galaxies grow, black holes develop as a result of 
galaxy mergers \citep{kauffmann00} and become active in
suppressing star formation in their host galaxies \citep{schawinski06}. 
Throughout this paper, we use the following set of cosmological 
parameters: $\Omega_0=0.27$, $\Omega_\Lambda=0.73$,
$\Omega_{\rm b}/\Omega_0=0.15$, $\sigma_8=0.77$, and $h=0.71$.

\subsection{Dynamical Friction and Gas Stripping}

Earlier implementations of the dynamical friction of 
satellite haloes were based on the \citet{chandra43} formula with 
a fixed satellite halo mass \citep[e.g.][see however \citealt{taylor04} for models with mass loss]{kauffmann99}.
Detailed numerical simulations, however, have found  that the Chandrasekhar 
formula systematically underestimates the merging timescales, 
especially for large mass ratios ($\mhost/\msat\gg1$) \citep[e.g.][]{boylan08}.

To delineate the evolution of satellite systems more precisely 
during halo mergers, we calculate the two-dimensional orbital motions and 
tidal mass-loss of the satellite halo.
We treat each subhalo as a point-like object orbiting in an 
isothermal host potential. Satellite galaxies embedded in a subhalo are 
initially placed at the virial radius of the host halo at the beginning of 
mergers, orbiting  with the circular velocity of the host halo 
\citep{benson05,khochfar06b}. 
Tangential and radial velocities of the infalling haloes are 
assigned as ($\eta V_c$, $\sqrt{1-\eta^2}V_c$), where $\eta$ is the circularity 
defined as the ratio of the orbit angular momentum to that of circular 
orbit with the same energy. We adopt the random circularity following 
\citet{lacey93}. 
The exerted dynamical friction is computed 
as \citep{bt98},
\begin{align}
\frac{\dd \vec{v}}{\dd t} = & - \frac{G\msat(t)}{r^2} \ln \Lambda \\ 
& \left(\frac{V_c}{v}\right)^2 \left\{ {\rm erf} \left(\frac{v}{V_c}\right)
 -  \frac{\sqrt{\pi}}{2} \left(\frac{v}{V_c}\right)
\exp \left[- \left( \frac{v}{V_c}\right)^2\right]\right\} \vec{e}_v ,
\end{align}
where $\Lambda$ is a Coulomb logarithm and we use the value of 
($1+\mhost/\msat$) following \citet{springel01}, $V_c$ is the 
circular velocity of the halo at the virial radius, $v$ is the orbital 
velocity of the satellite halo, $\vec{e}$ is the unit velocity vector 
and $r$ is the distance of the satellite from the centre of the host halo. 
Here the satellite mass \msat\ explicitly depends on time. 

To evaluate the satellite mass \msat\ during its orbit, 
we calculate the radius within which the motion of a particle is 
governed by the satellite. We assume that all satellite mass within 
the sphere of influence ($\rsoi$) is bound to the satellite, 
while the matter outside of $\rsoi$ is stripped.  
The sphere-of-influence radius (\rsoi) can be written as \citep{battin87}
\begin{align}
\rsoi \sim r \bigg[  \left( \frac{ \msat(t)}{\mhost(<r)}\right)^{-0.4}& (1+3 \cos^2\theta)^{0.1} \nonumber \\
& + 0.4\cos\theta \left(\frac{1+6 \cos^2\theta}{1+3\cos^2\theta}\right) \bigg]^{-1} 
\label{eqn1}
\end{align}
where $r$ is the separation between the central and satellite galaxies, 
 $\msat(t)$ is the total (baryon$+$dark matter) mass of the satellite halo, 
$\mhost(<r)$ is the total mass of the central halo within $r$, and 
$\theta$ is the angle between the line connecting the particle to the center of
mass of the satellite halo and the line connecting the centers of mass of
the satellite and the host haloes.
For large mass ratios ($\mhost/\msat \gg 1$), Eqn. \ref{eqn1} can be reduced to 
$\rsoi \sim r[\msat(t)/\mhost(<r)]^{2/5}$.
At each time step ($\delta t$), we assume that a fraction ($\delta t/t_{\rm dyn}$) 
of dark matter outside  \rsoi\ is stripped, where $t_{\rm dyn}$ is 
the dynamical time scale of the satellite halo. We assume that a galaxy merger 
takes place if the separation between the two haloes becomes smaller 
than the size of a central galaxy, which is computed from the observational 
relationship between galaxy mass and size \citep{shen03}. 
We assume the size of a galaxy to be twice the effective radius.
 By doing so, the merging time scale obtained from our simple analytic model 
 shows good agreement  with the numerical study of \citet{boylan08}  
 (within 20\% error). 

It is interesting to compare our simple model for mass loss with 
other studies. Our calculation gives a mass loss of approximately 40\% for the 
dark matter component during the first peri-centric passage, 
which is comparable to what is found in \citet{taylor04}. 
Note that our calculation does not include the effect of tidal heating. 
We have also compared our mass-loss curves with results 
obtained through N-body simulations using 
GADGET-2 (Yi et al. in prep), and have found qualitatively good agreement. 

As a satellite orbits in the host potential, the diffuse gas of the 
satellite will be  stripped due to the ram pressure exerted by 
the intracluster medium \citep{gunn72}. We model the ram pressure 
following the analytic formulation of 
\citet[][see also \citealt{font08}]{mccarthy08}. 
We first compute a radius $\rram$ beyond which ram pressure is able to strip material
\begin{equation}
\rho_{\rm host}(r) v_{\rm orb}^2(t) = \frac{2 G\msat(<\rram)\rho_{\rm gas}(\rram)}{\rram}, 
\end{equation}
where $\rho_{\rm host}$  is the density of the host halo gas, $v_{\rm orb}$ is the orbiting velocity of a satellite, and 
$\rho_{\rm gas}$ is the gas density of a satellite halo.
Similar to \rsoi, we remove a fraction 
($dt/t_{\rm cross}$) of diffuse gas outside \rram. However, if \rsoi\ is 
smaller than \rram, we take \rsoi\ as the  stripping radius. 
Using this analytic approach, we obtain halo gas loss of approximately 60\% 
after the first pericentre passage for 
$(\mhost/\msat,v_r/v_c, v_t/v_c)=(25, 0.9, 0.7)$, 
which is comparable to the results of \citet{mccarthy08}. 
It is worth mentioning that this gas stripping prescription 
is similar to \citet{guo10}, however the two models differ slightly 
in several manners. For example, they adopt a fixed orbiting velocity 
for satellite haloes, whereas our model follows the evolution of 
the orbital motion until the galaxy at the center of the halo 
merges with the central galaxy.

As the outer part of the satellite diffuse halo gas is stripped during its orbit, 
we assume that the remaining gas quickly restores an isothermal profile by 
redistributing the mass inside the stripping radius. Thus, the cooling rate 
becomes smaller as more diffuse gas is lost. Cold gas replenished by the 
satellite halo gas can subsequently form stars, and supernova explosions 
are assumed to blow cold gas out into the satellite diffuse halo. This 
assumption affects the star formation history of satellite galaxies, 
and we will discuss the implications in \S 4.

\subsection{Stellar Mass Loss}

A large fraction of stellar mass is believed to be released in
the form of cold gas and becomes available as a source of the next
generation stars.
Stellar evolution and mass loss strongly depend on stellar mass itself.
Low-mass stars evolve with a long characteristic timescale, whereas
high-mass stars are short-lived and end with dramatic mass loss processes.
This difference gives an important 
characteristic of stellar mass loss for each stellar population.
Since the mass spectrum of the population is continuous, stars evolve off 
the population either through supernova explosion or a planetary 
nebula process ${\it continually}$. 

Conventional SAM models often neglect this fact
because tracing stellar masses from the entire stellar subpopulations 
is a computationally expensive task. 
Instead, they generally calculate the fraction of ejecta for a given stellar
population and apply it to all other populations regardless of age. 
For example, when the Scalo initial mass function \citep{scalo86} is adopted, a 
stellar population will return roughly 30 percent of its mass during its 
evolution. Then, the models assume that 30 percent of newly formed stars 
quickly evolve off and constitute the galaxy's gas component. This may be a 
reasonable approximation for studies of the long-timescale star formation 
history of galaxies, but it is not as applicable to short-timescale histories, 
which are the main focus of our investigation.

In this study we develop an economic prescription whereby stellar 
mass loss can be approximated with reasonable accuracy. 
Since massive young stars evolve off  quickly and their remnant 
fraction in mass is low (i.e., a higher mass loss rate), a large mass-loss 
rate can be expected for galaxies with young populations. 
On the other hand, if the mean age of a galaxy 
is high, low-mass stars dominate the stellar 
mass loss, thereby lowering it. 
Therefore, one can anticipate the anti-correlation between the stellar 
mass-weighted age of a galaxy (\tm) and the 
specific stellar mass-loss rate  
($\dmloss \equiv dm_{\rm loss}/\dd t/\mgal$).

In order to obtain a relation between $\dmloss $ and \tm, we first compute the lifetime of stars 
with a broken-power law \citep{ferreras00}, which is obtained 
from the data of \citet{tinsley80} and \citet{schaller92}. 
\begin{equation}
\frac{\tau_m}{\rm Gyr} =
\left\{ \begin{array}{ll}
9.694 \left(m/M_{\odot}\right)^{-2.762} ~~~~~ m < 10 M_{\odot} \\
0.095 \left(m/M_{\odot}\right)^{-0.764} ~~~~~ m > 10 M_{\odot}.
\end{array} \right.
\end{equation}
At the end of their lifetimes, 
stars will release most of their mass, leaving small remnants such as 
white dwarf or black holes. The mass locked up in the 
remnant is approximated following \citet{ferreras00} as: 
\begin{equation}
\frac{w_m}{M_{\odot}} =
\left\{ \begin{array}{lll}
0.1 (m/M_{\odot}) + 0.45 ~~~~~~~~ m < 10 M_{\odot} \\
1.5 ~~~~~~~~~~~~~~~~~~~~~~~~~~~~ 10 \leq m < 25 M_{\odot} \\
0.61 (m/M_{\odot}) - 13.75 ~~~~~ m \geq 25 M_{\odot}.
\end{array} \right.
\end{equation}
Using this, we compute the stellar mass loss of galaxies with a variety 
of star formation histories, drawn from the fiducial SAM  without 
our new features. Assuming that newly formed stars at each redshift 
follow the  Scalo initial mass function, the specific stellar mass-loss rate 
is computed as a function of \tm\ following cosmic star formation 
histories. This \tdm\ relation will be used to estimate the 
approximate amount of stellar mass loss for a given simulation period 
($=\dmloss \mgal  \dd t $) in our SAM. This approximation is very 
reliable for ordinary star formation episodes, but it underestimates 
the mass loss when star formation rates are high (e.g., if there is a 
merger-driven starburst in the previous time step). In order to 
compensate for this shortcoming, we explicitly add the mass loss from 
the youngest stellar population in the model galaxy to the mass loss 
estimated from the \tdm\ relation. This treatment can be expressed as follows:
\begin{equation}
\delta m_{\rm loss} \simeq \left[ dm_{\rm loss}'  (t_m) ~ m_{\rm gal} + dm_{\rm loss}' (t_{ y}) ~ \delta m_{ y}\right] \delta t
\label{eqn3}
\end{equation}
where $t_m$ is the mass-weighted mean stellar age of the target galaxy, 
$t_{\rm y}$ is the mass-weighted mean age of its youngest populations,
 $\delta t$ is the integration timestep used in our SAM ($\approx$250 Myr 
 at $z=0$ and $\approx$80 Myr at $z=1$), and $\delta m_{\rm y}$ is the 
 mass of the stars formed since the last time step. 
Our modeling of non-instantaneous stellar mass loss is simpler 
 than the recent models of \citet{arrigoni10} or \citet{benson10}. 
 They fully consider the mass loss for the past time-steps, and include a 
 metallicity-dependent stellar yield \citep{portinari98,marigo01} 
 in the latter paper, which is neglected in this work.

\begin{figure}
\begin{center}
\includegraphics[width=8cm]{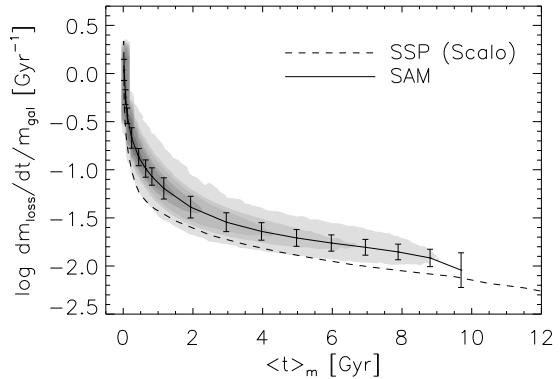}
% gemsv:~/SAM/Chemical/MassLoss/Large_run/Iter4/fit_mloss_con.pro
\caption{The relation between the stellar mass-weighted age of galaxies (\tm) 
and specific stellar mass-loss rate ($\dmloss=dm_{\rm loss}/\dd t/\mgal$).
Stellar mass-loss is estimated for galaxies with a variety 
of star formation histories, drawn from a semi-analytic model (gray shading). 
The mean and 1$\sigma$ error of this relation is shown as a solid line. 
The specific stellar mass-loss rate obtained from a simple stellar population 
with the Scalo IMF is shown as a dashed line. 
The \tdm\ relation is used to approximate the continuous stellar mass 
loss at each time step.  
} \label{fig:tdm}
\end{center}
\end{figure}

Fig. \ref{fig:tdm} shows the specific mass-loss rates 
of galaxies having a wide variety of star formation histories as grey 
contours and the \tdm\ relation as a solid line. The specific mass-loss 
rate of a simple stellar population is also shown as a dotted line.
Since subpopulations of a galaxy have different ages and younger 
populations yield more mass loss, the total \dmloss\ is usually 
greater than that of a simple stellar population.
As expected, the specific stellar mass-loss rate is  
a strong function of the  mass-weighted age of a galaxy. 
It is interesting to note that stellar mass loss rates are so small that only 
a population younger than $100 Myr$ can supply enough cold gas 
to make galaxies switch from passive to active, assuming a star formation 
efficiency of 2\% (Eq. \ref{ssfr_cut}).

A major source of uncertainty in our \tdm\ approximation is the fraction of 
stellar mass loss that contributes to the diffuse halo gas and cold disc gas. 
\citet{parriott08} show using two dimensional hydrodynamic simulations 
including radiative cooling that 25\% of the mass loss from red giants 
remains cool. On the other hand, most of the mass loss from the planetary 
nebula phase appears to remain warm or cool \citep{bregman09}.
Since the main contribution of the mass loss comes from 
stars with masses of  2 -- 8 $\msun$, whose final stages are 
planetary nebula, we assume that half of the mass loss 
goes into the  hot halo gas and the rest returns to the cold 
disc gas reservoir. 
The exact values of the fractions are unknown, but the models are not
very sensitive to such variations (within a factor of two).

\subsection{The effects of the new prescriptions}

\begin{figure}
\begin{center}
\includegraphics[width=8cm]{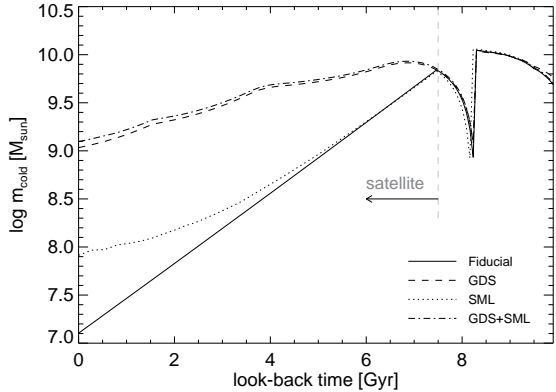}
% asosx100:~/Research/MassLoss/Paper/fig4/two_ex.pro
\caption{The impact of the different physical prescriptions on the cold gas 
evolution in a sample satellite galaxy. The vertical dashed line marks the 
moment where a central galaxy becomes a satellite galaxy through a halo 
merger. The fiducial model (solid line) assumes that the hot gas halo of the 
satellite is shock heated and instantly stripped away at the time of the merger. 
Note that the fiducial model does not take into account stellar mass loss.
The dotted line shows the model in which stellar mass loss contributes to the 
cold gas disc. The dashed line shows the model that includes the gradual 
diffuse gas stripping prescription but not the stellar mass loss. The dash-dot 
line shows our new model, which includes both features.  
The rapid drop in cold gas around 8 Gyr is 
due to a major merger that turned much of the cold gas into stars. }
 \label{fig:one_ex}
\end{center}
\end{figure}

In Fig. \ref{fig:one_ex} we show the impact of the two different physical 
prescriptions on the cold gas evolution for a $2\times 10^{10} \msun$ 
disc-dominated satellite galaxy. The satellite galaxy can retain a copious 
amount of cold gas from diffuse gas cooling if it maintains a hot gas halo 
for an extended period of time. This leads to a further growth in stellar mass 
in the satellite galaxy. It should be noted that the contribution from stellar 
mass loss is an order of magnitude smaller than the prolonged gas cooling 
in this specific example, and results in an order of magnitude increase in cold gas at $z=0$,
compared with our fiducial model where no recycling is used. 

\section{Results}

In this section, we investigate the effect of the two aforementioned 
physical processes on the recent star formation history of satellite 
galaxies. There are various ways to quantify recent star formation 
activity; we make use of the specific star formation rate (SSFR$\equiv$SFR/\mgal), 
obtained from UV and optical photometry.
UV lights are much more sensitive to current and
recent star formation than optical colours, and, compared with emission line 
diagnostics, they detect not only on-going but also recent (up to about one billion 
years) star formation.  
Empirical specific star formation rates are the values averaged over the 
past 100 Myr \citep{salim07} and model values are averaged over the past 
270 Myr. 

We consider a galaxy ``passive'' if the following condition is met:
\begin{equation}
\log {\rm SSFR} \leq -9 -0.2 \log (\mgal/h^{-2}\msun), 
\label{ssfr_cut}
\end{equation} 
We choose this threshold because GALEX ultraviolet detections and 
star formation rate measurements are robust above this level. 
This criterion reasonably divides galaxies into active and passive galaxies (see K09).

The observational sample used here is the same used in K09, 
which is drawn from the GALEX-SDSS matched sample constructed by 
\citet{salim07}. K09 also make use of the Group catalog \citep{yang07}
for estimates of the dark matter halo mass and galaxy stellar mass. 
For a more detailed description of the data, readers are referred to K09.

\subsection{Recent Star Formation And Morphology}

\begin{figure}
\begin{center}
\includegraphics[width=8cm]{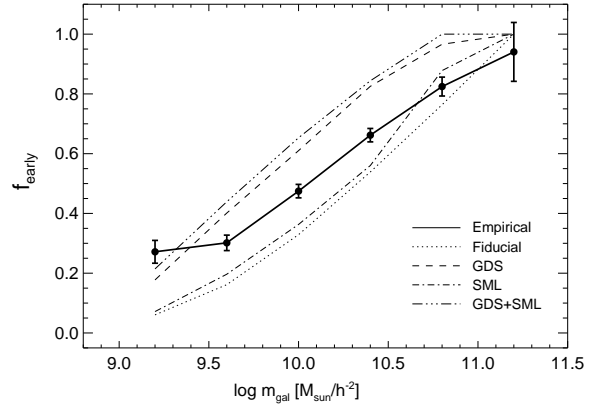}
% gem:~/SAM/Analysis/Volume/Mstar_vs_fpas/Observation/draw_morph.pro -> fmorph_sat.dat
% gem:~/SAM/Analysis/Volume/Mstar_vs_fpas/draw_morph_ysam.pro
\caption{The fraction of early-type galaxies (\fearly) as a function of galaxy 
stellar mass. We assume that early-type galaxies have r-band concentration 
indices greater than 2.6 in the observational sample.
} \label{fig:morph}
\end{center}
\end{figure}

\citet{kimm09} inspected star formation rates with respect to galaxy mass
and halo mass, but did not consider galaxy morphology.
It is important to check whether the satellite over-quenching problem 
is present in all morphological types of galaxies. In order to examine this, 
we divide the observational data into early and late types using the concentration index
in the r-band by requiring that early types have $C_r=r_{90}/r_{50}>2.6$ 
\citep{strateva01,shimasaku01}.
Model galaxies are classified based on stellar mass-weighted bulge-to-total ($B/T$) ratio. 
We assume that early types have a $B/T$ ratio of greater than 0.4 (Fig. \ref{fig:morph}),
which is motivated by observational studies \citep[e.g.][]{simien86},
where the division occurs around B/T$=$0.4.

\begin{figure}
\begin{center}
\includegraphics[width=8cm]{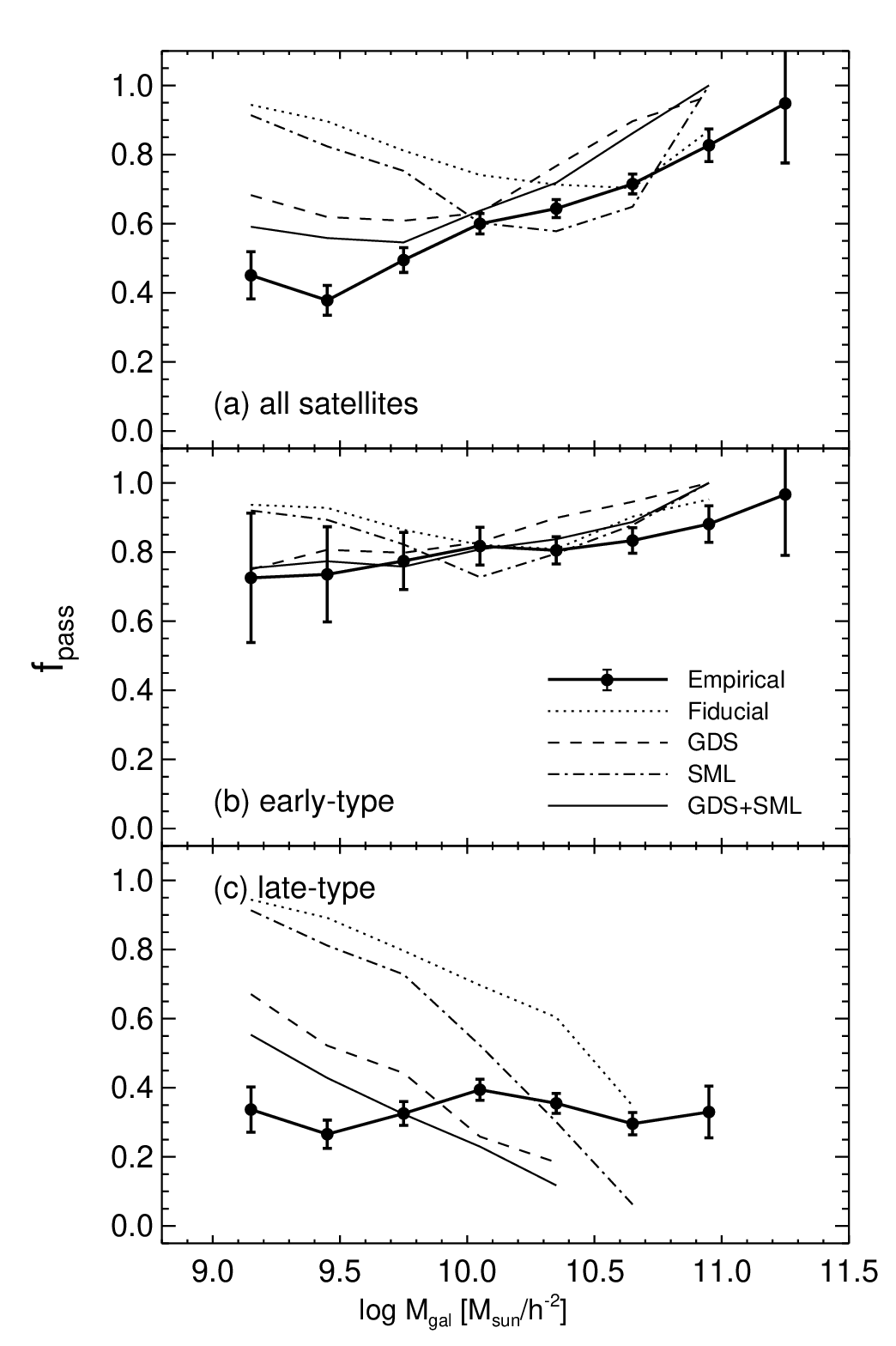}
% gem:~/SAM/Analysis/Volume/Mstar_vs_fpas/mstar_vs_fpass.pro
\caption{The fraction of passive galaxies (\fpass) as a function of 
galaxy stellar mass. The top panel shows the \fpass\ for all satellite galaxies. 
Early-type (middle) and late-type (bottom) galaxies are also shown separately.
The empirical data is indicated by solid lines, while our fiducial model (old), 
the model with gradual diffuse gas stripping (GDS), the model with stellar mass 
loss (SML), and the model with both GDS and SML are shown by black dotted, 
green dashed, orange dotted-dashed, and red solid lines, respectively 
(see chart key). Our original semi-analytic model overproduces passive late 
types. Note that late-type galaxies become more active (consistent with observations) when GDS and SML are take into account.} 
\label{fig:fpasmgal}
\end{center}
\end{figure}

Fig. \ref{fig:fpasmgal} shows the fraction of passive satellites 
($\fpass$) for different morphologies. The observational data (Fig. \ref{fig:fpasmgal}-(a))
show a strong mass dependence of the passive galaxy
fraction. Interestingly, the mass dependence of \fpass\ virtually 
disappears when focusing on individual morphological classes (Figs. \ref{fig:fpasmgal}-(b) and (c)), which implies that the overall mass dependence is a result of a combination of scaling relations 
among galaxy mass, type, and star formation rates (Fig. \ref{fig:fpasmgal}) \citep[e.g.][]{park07}.

The morphological mixture as a function of galaxy mass is 
reasonably reproduced by our fiducial SAM (Fig. \ref{fig:fpasmgal}). 
The passive galaxy fractions for early types are reasonably reproduced 
by most models as long as they include AGN feedback \citep{schawinski06,kimm09}. 
All models present here include AGN feedback.
However, most late-type model galaxies do not match observations, 
as they generally predict that at least 80\% of late types must be passive
 (Fig. \ref{fig:fpasmgal}-(c)).
As discussed previously, late-type galaxies cannot maintain the observed level of star formation, most likely because of the lack of cold gas.

Fig. \ref{fig:fpasmgal}-(c) shows how our new prescriptions change the 
model outputs against the observed late-type galaxies. Model galaxies 
become progressively more active when stellar mass loss and gradual 
gas stripping are considered. The inclusion of stellar mass loss alone still 
produces too many small passive spirals. This is because the amount of 
stellar mass loss released from old stellar populations in small galaxies is 
too little to fuel star formation. When both gradual diffuse gas stripping 
and continuous stellar mass loss are adopted, the passive galaxy fraction 
becomes closer to the data. As an indication of the sensitivity of the 
observational fraction of passive galaxies to the actual concentration index 
cut $C_r$, we choose a cut of $C_r=2.4$ as an example.  
This cut is more conservative in the sense that it provides a cleaner late-type sample. We 
find that the more strict cut results in a lower passive galaxy fraction 
($\fpass\sim0.1 -0.3$), indicating possible contamination from early-type 
galaxies within the sample selected using the less conservative cut. 
It should be noted that our models predict a strong dependence of the passive 
fraction on  galaxy mass for late type satellites, which is not observed. 
This indicates that the evolution of cold gas in late type satellites is not 
yet correctly captured by the model.

The inclusion of stellar mass loss and gradual diffuse gas stripping affects 
the morphology of model galaxies in a counterintuitive way. Fig.~\ref{fig:morph} 
shows that such considerations raise the early-type galaxy fraction. 
This may sound counterintuitive, because those processes generally 
enhance the star formation in the disc. However, the morphology 
(the bulge-to-total ratio) of a satellite galaxy rarely changes due to these 
considerations. The increase in the disc mass due to prolonged star formation 
can be large enough to make a galaxy ``active'', but is too small to change the 
galaxy morphology. On the contrary, an increase in the stellar disc occurring 
in the central galaxy leads to increases in the bulge mass of the central galaxy 
whenever it merges with another galaxy, according to the current semi-analytic 
prescription. This has the net effect of increasing the number of galaxies with 
a higher value of bulge-to-total ratio.

Our criterion for passive galaxies is rather conservative, and it is interesting 
to note that relaxing this criterion to include intermediate colors alleviates 
the discrepancy between model and observation. This implies our model produces 
too many intermediate-color late types \citep[see also][]{weinmann10}.

\subsection{Environmental Dependence}

\begin{figure*}
\begin{center}
\includegraphics[width=16cm]{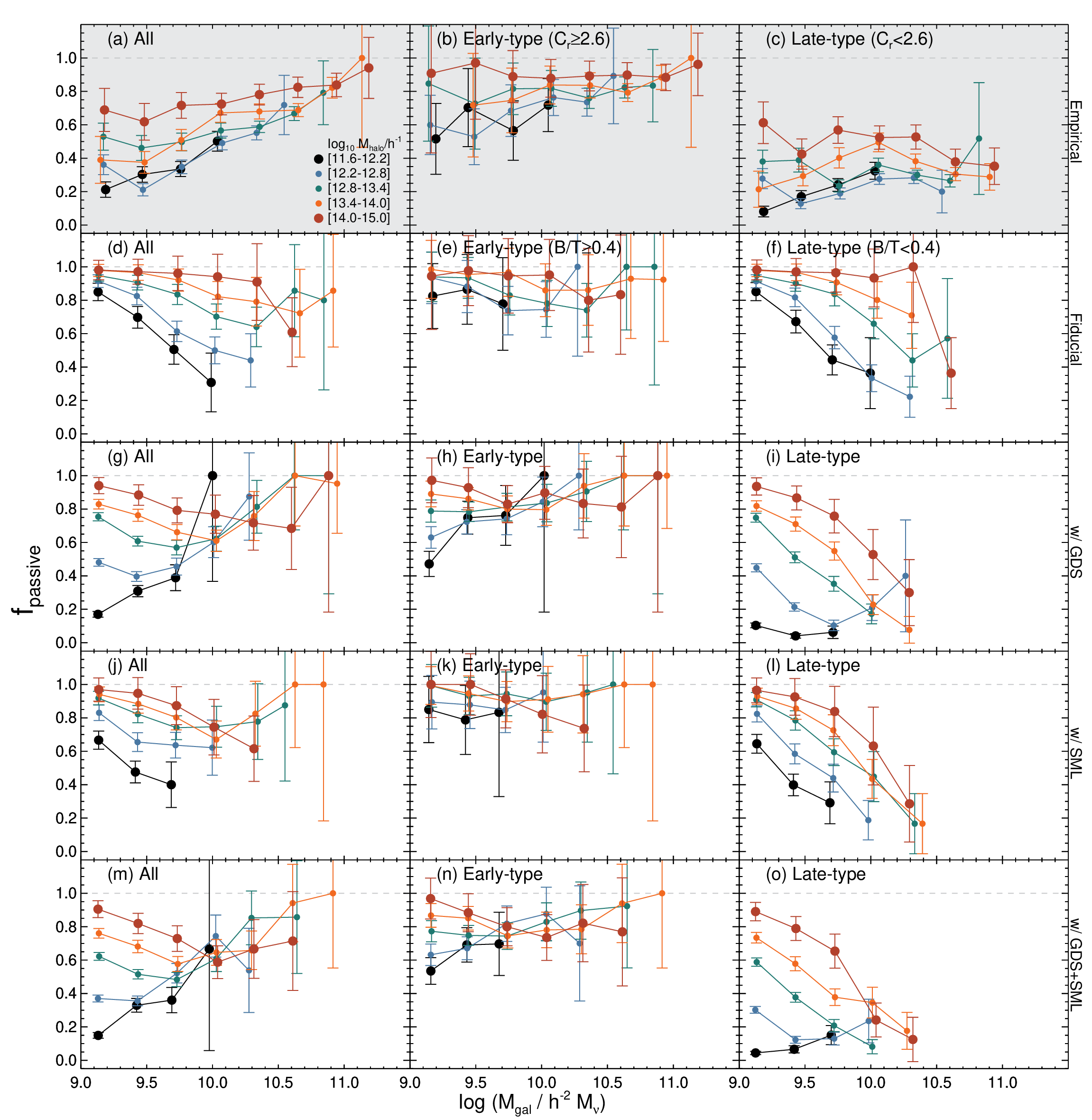}
% gem:~/SAM/Analysis/Volume/Mstar_vs_fpas/Model/dplot6.pro
\caption{The passive fraction of satellite galaxies (\fpass) as a function of 
galaxy stellar mass, for different halo mass bins, as indicated by 
different colours and symbol sizes (see plot legend). Top panels display 
the observed values. The theoretical models with different prescriptions 
are shown in the subsequent rows. We present the results for different 
morphologies, as indicated in the plot. The SAM with gradual diffuse gas 
stripping (GDS) and stellar mass loss (SML)  shows  improved agreement with the data,
although small galaxies in massive haloes are admittedly still too passive.
} \label{fig:fpas}
\end{center}
\end{figure*}

The observed fractions of passive satellite galaxies for different 
environments (halo mass) and morphologies are shown in Fig. \ref{fig:fpas} 
(top row). The estimates of dark matter halo masses for observed 
galaxies are adopted from the Group catalog constructed by \citet{yang07}. 
The colour-codings represent different halo masses. 
As can be seen, both early and late types show an environmental dependence 
in that star formation activity in larger clusters is suppressed. 
Again, when the morphology is fixed, the dependence of \fpass\ on galaxy mass 
and halo mass is weak.

The environmental dependence can also be found in our original (fiducial) 
SAM (the second row from the top) where the diffuse halo gas of satellites 
is instantly shock-heated at the beginning of halo mergers. This can be 
understood by noting that satellite galaxies in more massive halos merge 
with their host halo earlier  and use up their available cold gas without any  
replenishment. This mechanism is even clearer for late types, because early-type 
satellites are more strongly influenced by AGN feedback in our models 
\citep{schawinski06}.  Note that more than 80\% of the small late types 
in the smallest haloes (black line) are predicted to be passive in this version, 
which is inconsistent with observation.

Models with gradual gas stripping (third row from the top) shows a 
better match to the data. The extended cold gas 
supply preferentially lowers the passive galaxy fraction of massive galaxies. 
This is because the gravitational restoring force of such systems is usually 
strong enough to retain a large amount of halo gas. 
Since the ram pressure becomes less effective if the mass ratios  
between the host and the satellite halo are comparable, galaxies in less massive 
haloes are more likely to be active, for a given galaxy mass. 
However, this model still predicts that a substantial fraction of small late types in massive haloes are passive. Hence, we argue that gradual gas stripping alone cannot reproduce the observed level of recent star formation activity. 

The effect of stellar mass loss on recent star formation history appears to
be smaller than that of gradual diffuse gas stripping (fourth row
from the top). Most galaxies in large haloes are predicted to be passive, which 
is inconsistent with observation. It seems that stellar mass 
loss alone cannot maintain the observed level of star formation of late types 
in cluster environments. This is still true even if we arbitrarily allow 
100\% of the stellar mass loss to contribute to the cold phase gas.
Lastly, the model with both gradual diffuse gas stripping {\em and} stellar 
mass loss shows a further improvement in matching the observed fraction 
of passive small satellites  (bottom row). Since cold gas supplies are established 
through two channels, the overall passive galaxy fraction is diminished.
This model slightly underproduces the massive and passive late types. 
When the total sample is considered (left column), the models appear to reproduce
the observed dependence of passive galaxy fraction on galaxy stellar mass.
Based on our simple exercise using gradual diffuse gas stripping and
stellar mass loss, we conclude that both contribute to the recent star formation
history of satellite galaxies.

\section{Limitations}
We have shown that allowing a diffuse gas reservoir and continuous 
stellar mass loss for satellite galaxies can further supply cold gas, and hence 
may increase the fraction of actively star-forming spirals in group environments.  However, we note that the amount of the supplied gas that is actually used to 
form stars depends on the mechanism of the supernova feedback. 

As an extreme case, if supernovae do not strongly blow out cold gas from 
discs, more would be available for star formation, and the fraction of passive 
spirals would decrease. Additionally, satellite galaxies would form stars 
for a long period of time \citep[e.g][]{kennicutt98} even if diffuse gas 
reservoir were depleted. 
On the other hand, if supernova explosion blow most of the cold gas away 
from the satellite halo, there would be less hot gas remaining. 
Since the cooling rate is comparable to the blowout rate by supernovae in 
small subhaloes, where the disagreement in the passive fraction is the most 
notable, the amount of cold gas would drop rapidly.  Then, most of the small 
galaxies that have already moved around their central galaxy several times are
more likely to be passive at $z=0$ even if they may have their own hot gas halo. 
This may suggest that satellite over-quenching 
might not be solely due to inadequate treatment of environmental effects, 
but may also relate to how one models supernova explosions.

In this study, we have assumed that supernova explosion cannot blow cold 
gas away but instead transforms it to hot diffuse gas. Clearly, this is an 
oversimplification, and further investigation is needed on this subject. 
It is interesting to note, however, that relaxing the assumption so that the 
reheated gas can escape the satellite potential would exacerbate the mismatch 
in the passive fraction for small disc-dominated satellite galaxies. 
It is also possible that the supernova feedback is ineffective in 
transforming cold gas into hot gas in the galaxies we discussed \citep[e.g][]{maclow99,dubois08}.

\section{Conclusions and Discussion}

We have investigated the recent star formation history of satellite galaxies 
by comparing semi-analytic models of galaxy formation with empirical data 
drawn from SDSS and GALEX. 
Based on the star formation rate measurements derived from multiband 
photometry \citep{salim07}, we first divided galaxies into active 
 and passive types, and computed the fraction of passive galaxies 
(\fpass) as a function of galaxy stellar mass. As already shown in  
previous studies \citep{weinmann06,baldry06,kimm09}, satellite galaxies in 
theoretical models cannot reproduce the observed level of recent star formation 
activity. The satellite over-quenching problem is generally attributed to the strong 
strangulation applied in most semi-analytic models in which diffuse gas of the 
satellite system is instantly shock-heated from the system at the beginning of 
halo mergers \citep[e.g.][]{white91}. 

In the hope of improving the situation, we implement {\em gradual diffuse gas 
stripping} and {\em stellar mass loss} into our semi-analytic code.
For detailed tests, we divide the sample by morphology. We have also 
introduced an  approximation for the contribution to the ISM from stellar 
mass loss. Our results can be summarised as follows.
\begin{itemize}
\item The over-quenched satellite galaxies in SAMs are mostly late-type. 
\item The models with gradual diffuse gas stripping resolve much of the
satellite over-quenching problem. However, they cannot account for the 
significant fraction of actively star-forming small late-type galaxies 
observed in massive haloes. 
\item Stellar mass loss is not a dominant source of cold gas in most satellite galaxies
but enhances residual star formation.
\item Our new models incorporating both effects show a significantly-improved 
match to the observed data. However, they still suggest that the majority of cluster late types are
passive.
\end{itemize}
Recent SAMs including ours have shown steady progress in matching the observed data, 
but are missing key ingredients. For example, feedback from supernova 
explosions affects the recent star formation history in satellite galaxies by 
regulating the remaining cold disc gas but is still poorly constrained (\S 4).
 
One might wonder whether or not simply increasing the contribution of 
stellar mass loss to the cold ISM can reproduce the passive 
late-type fraction. We have found that such an approach still shows a lack of 
actively star-forming late types. \citet{kaviraj07} reached a similar 
conclusion when trying to explain early-type galaxies with residual star formation  \citep{yi05}.

Recently, \citet{tonnesen09} demonstrated that a {\em weak} ram pressure 
does not only remove cold gas in satellites but also may enhance star 
formation by compressing the gas component. Provided that the cooling of cold 
gas onto a galaxy disc is suppressed during the interaction between the host
and satellite haloes and the satellites are affected by the weak ram pressure 
for a long time, this may increase \fpass\ by consuming a significant fraction 
of cold gas. Yet, it is still unclear how gas cooling takes place 
in massive satellite systems during halo merging. 
Since cooling also relies on the density, if ram pressure compresses both cold 
and diffuse gas, cooling could also occur more efficiently. 
If this is the case, the efficient cold gas consumption might have little  
effect on \fpass. Detailed numerical simulations are necessary to better understand 
the supply of cold gas during interactions.

An accurate determination of initial orbits may have an impact on \fpass.
Based on numerical simulations at $z\sim0$ 
\citep{benson05,zentner05,khochfar06b},  we have assumed that satellite 
galaxies have random circularities at the time of halo mergers over the entire 
cosmic history. If radial orbits are more common at higher redshifts 
\citep[e.g.][]{wetzel10,dekel09}, gas stripping due to ram pressure would be more 
effective because satellites penetrate dense intra-cluster media more frequently 
on radial orbits. Motivated by this idea, we performed a simple experiment:
we assigned an eccentricity of 1 to the haloes that experience halo merger 
at $z\geq 1$. In later mergers, orbital parameters were randomly chosen in 
terms of circularity, as done in the fiducial model. 
Since massive galaxies are likely to orbit several times while smaller galaxies 
have passed the pericentre less, the influence of ram pressure on radial orbits 
is more notable in massive satellite galaxies. As a result, the negative 
correlation of \fpass\ with \mgal\ shown by our new late-type galaxy models in 
massive haloes gets diminished, making the models match the data slightly better.
However, the reliability of our exercise depends on the validity of
the demarcation redshift and actual eccentricity distribution assumed.

We also note that ignorance of the tidal disruption could have an impact on 
the passive galaxy fraction. \citet{taylor04} and \citet{zentner03} showed 
that subhaloes lose 30--40 per cent of their mass per pericentric passage. 
This implies subhaloes are dissolved into host haloes after several orbits 
unless they merge with centrals. 
On this basis, \citet{somerville08} implemented a prescription that
satellites are disrupted when the subhaloes lose $\sim90\%$ of their 
initial mass. By doing so, they reproduced the luminosity and  
the radial distribution of Milky Way satellites \citep{maccio10}. 
\citet{henriques08} also present good matches to galaxy colours
using a simple assumption that satellites which are not associated with 
subhaloes in the Millennium dark matter simulation \citep{springel05} 
are already disrupted by the tide in their environment.
Yet, as discussed in K09, the \citet{somerville08} models still show a
lack of actively star forming small satellites, implying that ignorance 
of tidal disruption is not the primary cause of red and dead small 
(late-type) satellites. Nevertheless, since it is small red and dead 
galaxies that are preferentially disrupted by tidal forces, the \citet{somerville08} 
models exhibit a slightly better agreement with empirical data than other 
semi-analytic models without tidal disruption (see K09 for details). In this regard, the passive 
galaxy fraction in the low-mass regime could be even smaller, 
resulting in a weaker dependence of the passive fraction on galaxy mass.

Recent semi-analytic models often adopt AGN feedback to prevent hot gas
cooling in massive galaxies \citep[i.e.][]{croton06}.
The effect has been particularly important for ``central'' galaxies that are fuelled 
by ongoing cooling, in contrast to satellite galaxies that cannot retain  
their hot halo gas in most SAMs. However, the strong strangulation 
appears problematic in that even late types suffer from the lack of cold gas. 
For that reason, we demonstrate that external as well as internal gas supply 
are necessary to match the observed fraction of passive galaxies. Allowing the 
external supply has an interesting implication. Since the gas retained in the hot 
halo could funnel into black holes \citep[e.g][]{croton06}, 
AGN feedback is likely triggered, and possibly suppresses star formation activity 
in satellite galaxies with supermassive black holes.  On the other hand, the effect 
of AGN feedback may be negligible in late-type galaxies because their black 
holes are not massive enough to release a large amount of energy. Instead, 
the evolution of late-type galaxies in clusters or groups may be more likely 
driven by the combination of stellar mass loss and environmental effects such 
as tidal and/or ram pressure stripping. We have indeed shown to first order that 
both mechanisms could reproduce observed levels of recent star formation, 
but it is still unresolved how late-type satellite galaxies show similar fractions 
of passive galaxies over various galaxy stellar masses for a given halo. 
Explaining these observations with realistic assumptions is an  
important challenge, especially since most of the galaxies in the universe are satellites.

\section*{Acknowledgments}
This work was supported by the Korean government through the Korea Research 
Foundation Grant (KRF-C00156) 
and the Korea Science and Engineering Foundation grant (No. 20090078756).


\begin{thebibliography}{99}
\bibitem[\protect\citeauthoryear{Abadi, Moore, \& Bower}{1999}]{abadi99} 
Abadi M. G., Moore B., Bower R. G., 1999, MNRAS, 308, 947
\bibitem[\protect\citeauthoryear{Arrigoni et al}{2010}]{arrigoni10} 
Arrigoni M., Trager S.~C., Somerville R.~S., \& Gibson B.~K. 2010, MNRAS, 402, 173 
\bibitem[\protect\citeauthoryear{Athanassoula}{2008}]{athanassoula08} 
Athanassoula E., 2008, MNRAS, 390, L69 
\bibitem[\protect\citeauthoryear{Baldry et al.}{2006}]{baldry06}
Baldry I. K., Balogh M. L., Bower R. G., Glazebrook K., Nichol R. C., 
Bamford S. P., Budavari T., 2006, MNRAS, 373, 469 
\bibitem[\protect\citeauthoryear{Battin}{1987}]{battin87}
Battin R. H., 1987, An Introduction to Mathematics \& Methods of Astrodynamics, AIAA, New York 
\bibitem[\protect\citeauthoryear{Bekki, Couch, \& Shioya}{2002}]{bekki02}
Bekki K., Couch W. J., Shioya Y., 2002, ApJ, 577, 651
\bibitem[\protect\citeauthoryear{Benson}{2005}]{benson05}
Benson A. J., 2005, MNRAS, 358, 551
\bibitem[\protect\citeauthoryear{Benson \& Bower}{2010}]{benson10}
Benson A.~J., \& Bower R. 2010, MNRAS, 405, 1573  
\bibitem[\protect\citeauthoryear{Binney}{1977}]{binney77}
Binney J., 1977, ApJ, 215, 483 
\bibitem[\protect\citeauthoryear{Binney \& Tremaine}{1998}]{bt98}
Binney J., Merriﬁeld M., 1998, Galactic Astronomy. Princeton Univ. Press Princeton
\bibitem[\protect\citeauthoryear{Boylan-Kolchin, Ma, \& Quataert}{2008}]{boylan08}
Boylan-Kolchin M., Ma C. -P., Quataert E., 2008, MNRAS, 383, 93
\bibitem[\protect\citeauthoryear{Bower et al.}{2006}]{bower06}
Bower R. G. et al., 2006, MNRAS, 370, 645
\bibitem[\protect\citeauthoryear{Bregman \& Parriott}{2009}]{bregman09}
Bregman J. N., \& Parriott J. R., 2009, ApJ, 699, 923 
\bibitem[\protect\citeauthoryear{Chandrasekhar}{1943}]{chandra43}
Chandrasekhar S., 1943, ApJ, 97, 255
\bibitem[\protect\citeauthoryear{Chung et al.}{2007}]{chung07}
Chung A., van Gorkom J. H., Kenney J. D. P., Vollmer B., 2007, ApJ, 659, L115
\bibitem[\protect\citeauthoryear{Cox et al.}{2008}]{cox08} 
Cox T. J., Jonsson P., Somerville  R. S., Primack J. R., \& Dekel A., 2008, MNRAS, 384, 386 
\bibitem[\protect\citeauthoryear{Croton et al.}{2006}]{croton06}
Croton D. et al., 2006, MNRAS, 365, 11
\bibitem[\protect\citeauthoryear{Dekel et al.}{2009}]{dekel09}
Dekel, A., et al.\ 2009, Nature, 457, 451 
\bibitem[\protect\citeauthoryear{Dressler}{1980}]{dressler80} 
Dressler A., 1980, ApJ, 236, 351 
\bibitem[\protect\citeauthoryear{Dubois \& Teyssier}{2008}]{dubois08} 
Dubois Y., Teyssier R., 2008, A\&A, 477, 79 
\bibitem[\protect\citeauthoryear{Ferreras \& Silk}{2000}]{ferreras00} 
Ferreras I., \& Silk J., 2000, ApJ, 532, 193 
\bibitem[\protect\citeauthoryear{Fontanot et al.}{2009}]{fontanot09} 
Fontanot F., De Lucia G., Monaco P., Somerville R. S., \& Santini P., 2009, MNRAS, 397, 1776 
\bibitem[\protect\citeauthoryear{Font et al.}{2008}]{font08} 
Font A. S. et al., 2008, MNRAS, 389, 1619  
\bibitem[\protect\citeauthoryear{Gilbank \& Balogh}{2008}]{gilbank08} 
Gilbank D.~G., \& Balogh M.~L., 2008, MNRAS, 385, L116 
\bibitem[\protect\citeauthoryear{Gisler}{1978}]{gisler78} 
Gisler G. R. 1978, MNRAS, 183, 633 
\bibitem[\protect\citeauthoryear{Gunn \& Gott}{1972}]{gunn72} 
Gunn J. E., Gott, J. R., 1972, ApJ, 176, 1
\bibitem[\protect\citeauthoryear{Guo et al.}{2010}]{guo10} 
Guo Q., et al. 2010, arXiv:1006.0106 
\bibitem[\protect\citeauthoryear{Jeltema et al.}{2007}]{jeltema07}
Jeltema T. E., Mulchaey J. S., Lubin L. M., \& Fassnacht C. D., 2007, ApJ, 658, 865 
\bibitem[\protect\citeauthoryear{Hatton et al.}{2003}]{hatton03} 
Hatton S., Devriendt J. E. G., Ninin S., Bouchet F. R., Guiderdoni B., Vibert D., 2003, MNRAS, 343, 75
\bibitem[\protect\citeauthoryear{Henriques, Bertone \& Thomas}{2008}]{henriques08} 
Henriques B. M., Bertone S., \& Thomas P. A., 2008, MNRAS, 383, 1649 
\bibitem[\protect\citeauthoryear{Kang \& van den Bosch}{2008}]{kang08} 
Kang X., van den Bosch F. C., 2008, ApJ, 676L, 101
\bibitem[\protect\citeauthoryear{Kauffmann et al.}{1999}]{kauffmann99} 
Kauffmann G., Colberg J.~M., Diaferio A., White S.~D.~M., 1999, MNRAS, 303, 188 
\bibitem[\protect\citeauthoryear{Kauffmann \& Haehnelt}{2000}]{kauffmann00} 
Kauffmann G., \& Haehnelt M., 2000, MNRAS, 311, 576 
\bibitem[\protect\citeauthoryear{Kauffmann et al.}{2004}]{kauffmann04} 
Kauffmann G., White S. D. M., Heckman T. M., M{\'e}nard B., Brinchmann J., Charlot S., Tremonti C., Brinkmann J., 2004, MNRAS, 353, 713 
\bibitem[\protect\citeauthoryear{Kaviraj et al.}{2007}]{kaviraj07} 
Kaviraj S. et al., 2007, ApJS, 173, 619
\bibitem[\protect\citeauthoryear{Kawata \& Mulchaey}{2008}]{kawata08} 
Kawata D., Mulchaey J. S., 2008, ApJ, 672L, 103
\bibitem[\protect\citeauthoryear{Kennicutt}{1998}]{kennicutt98} 
Kennicutt R.~C., Jr., 1998, ApJ, 498, 541 
\bibitem[\protect\citeauthoryear{Khochfar \& Burkert}{2005}]{khochfar05} 
Khochfar S., \& Burkert A., 2005, MNRAS, 359, 1379
\bibitem[\protect\citeauthoryear{Khochfar \& Silk}{2006}]{khochfar06} 
Khochfar S., \& Silk J., 2006, MNRAS, 370, 902 
\bibitem[\protect\citeauthoryear{Khochfar \& Burkert}{2006}]{khochfar06b} 
Khochfar S., \& Burkert A., 2006, A\&A, 445, 403 
\bibitem[\protect\citeauthoryear{Khochfar \& Ostriker}{2008}]{khochfar08} 
Khochfar S., \& Ostriker J., 2008, MNRAS, 680, 54
\bibitem[\protect\citeauthoryear{Kimm et al.}{2009}]{kimm09}
Kimm T. et al. 2009, MNRAS, 393, 1131
\bibitem[\protect\citeauthoryear{Lacey \& Cole}{1993}]{lacey93} 
Lacey C., \& Cole S. 1993, MNRAS, 262, 627 
\bibitem[\protect\citeauthoryear{Larson, Tinsley, \& Caldwell}{1980}]{larson80} 
Larson R. B., Tinsely B. M., Caldwell C. N., 1980, ApJ, 237, 692
\bibitem[\protect\citeauthoryear{Macci{\`o} et al.}{2010}]{maccio10} 
Macci{\`o} A.~V., Kang X., Fontanot F., Somerville R.~S., Koposov S., Monaco P., 2010, MNRAS, 402, 1995 
\bibitem[\protect\citeauthoryear{Mac Low \& Ferrara}{1999}]{maclow99} 
Mac Low M.-M., Ferrara A., 1999, ApJ, 513, 142 
\bibitem[\protect\citeauthoryear{Marigo}{2001}]{marigo01} 
Marigo P. 2001, A\&A, 370, 194 
\bibitem[\protect\citeauthoryear{McCarthy et al.}{2008}]{mccarthy08} 
McCarthy I. G., Frenk C. S., Font A. S., Lacey C. G., Bower R. G., Mitchell N. L., Balogh M. L., Theuns T., 2008, MNRAS, 383, 593
\bibitem[\protect\citeauthoryear{Park et al.}{2007}]{park07} 
Park, C., Choi, Y.-Y., Vogeley, M.~S., Gott, J.~R., III, \& Blanton, M.~R.\ 2007, ApJ, 658, 898 
\bibitem[\protect\citeauthoryear{Parriott \& Bregman}{2008}]{parriott08} 
Parriott J. R., \& Bregman J. N., 2008, ApJ, 681, 1215 
\bibitem[\protect\citeauthoryear{Portinari, Chiosi, \& Bressan}{1998}]{portinari98} 
Portinari L., Chiosi C., \& Bressan A. 1998, A\&A, 334, 505 
\bibitem[\protect\citeauthoryear{Quilis, Moore, \& Bower}{2000}]{quilis00} 
Quilis V., Moore B., Bower R., 2000, Science, 288, 1617
\bibitem[\protect\citeauthoryear{Rees \& Ostriker}{1977}]{rees77} 
Rees M. J., \& Ostriker J. P., 1977, MNRAS, 179, 541 
\bibitem[\protect\citeauthoryear{Salim et al.}{2007}]{salim07} 
Salim S. et al., 2007, ApJS, 173, 267
\bibitem[\protect\citeauthoryear{Schawinski et al.}{2006}]{schawinski06} 
Schawinski K. et al., 2006, Nature, 442, 888
\bibitem[\protect\citeauthoryear{Schaller et al.}{1992}]{schaller92} 
Schaller G., Schaerer D., Meynet G., \& Maeder A., 1992, A\&AS, 96, 269 
\bibitem[\protect\citeauthoryear{Scalo}{1986}]{scalo86} 
Scalo J. M., 1986, Fundamentals of Cosmic Physics, 11, 1 
\bibitem[\protect\citeauthoryear{Shen et al.}{2003}]{shen03} 
Shen S., Mo H. J., White S. D. M., Blanton M. R., Kauffmann G., Voges W., Brinkmann J., \& Csabai I.,  2003, MNRAS, 343, 978
\bibitem[\protect\citeauthoryear{Shimasaku et al.}{2001}]{shimasaku01} 
Shimasaku K. et al., 2001, AJ, 122, 1238
\bibitem[\protect\citeauthoryear{Silk}{1977}]{silk77} 
Silk J., 1977, ApJ, 211, 638
\bibitem[\protect\citeauthoryear{Simien \& de Vaucouleurs}{1986}]{simien86} 
Simien F., \& de Vaucouleurs G., 1986, ApJ, 302, 564
\bibitem[\protect\citeauthoryear{Somerville \& Kolatt}{1999}]{somerville99} 
Somerville R. S., Kolatt T. S., 1999, MNRAS, 305, 1
\bibitem[\protect\citeauthoryear{Somerville et al.}{2008}]{somerville08} 
Somerville R. S., Hopkins P. F., Cox T. J., Robertson B. E., Hernquist L., 2008, MNRAS, 391, 481
\bibitem[\protect\citeauthoryear{Springel et al.}{2001}]{springel01} 
Springel V., White S. D. M., Tormen G., Kauffmann G., 2001, MNRAS, 328, 726
\bibitem[\protect\citeauthoryear{Springel et al.}{2005}]{springel05} 
Springel V. et al., 2005, Nature, 435, 629
\bibitem[\protect\citeauthoryear{Strateva et al.}{2001}]{strateva01} 
Strateva I. et al., 2001, AJ, 122, 1861
\bibitem[\protect\citeauthoryear{Sun et al. }{2007}]{sun07} 
Sun M., Jones C., Forman W., Vikhlinin A., Donahue M., \& Voit M., 2007, ApJ, 657, 197 
\bibitem[\protect\citeauthoryear{Taylor \& Babul}{2004}]{taylor04} 
Taylor J. E., Babul A., 2004, MNRAS, 348, 811
\bibitem[\protect\citeauthoryear{Tinsley}{1980}]{tinsley80} 
Tinsley B. M., 1980, Fundamentals of Cosmic Physics, 5, 287 
\bibitem[\protect\citeauthoryear{Tonnesen \& Bryan}{2009}]{tonnesen09} 
Tonnesen S., \& Bryan G. L., 2009, ApJ, 694, 789
\bibitem[\protect\citeauthoryear{van den Bosch}{2008}]{vdb08} 
van den Bosch F.~C., Aquino D., Yang X., Mo H.~J., Pasquali A., McIntosh  D.~H., Weinmann S.~M., \& Kang X.,  2008, MNRAS, 387, 79 
\bibitem[\protect\citeauthoryear{Weinmann et al.}{2006}]{weinmann06} 
Weinmann S. M. et al., 2006b, MNRAS, 372, 1161
\bibitem[\protect\citeauthoryear{Weinmann et al.}{2010}]{weinmann10}
Weinmann S. M., Kauffmann G., von der Linden A., \& De Lucia G.,  2010, MNRAS, 406, 2249  
\bibitem[\protect\citeauthoryear{Wetzel}{2010}]{wetzel10}
Wetzel A.~R. 2010, arXiv:1001.4792 
\bibitem[\protect\citeauthoryear{White \& Frenk}{1991}]{white91} 
White, S.~D.~M., \& Frenk, C.~S.\ 1991, ApJ, 379, 52 
\bibitem[\protect\citeauthoryear{Yagi et al.}{2010}]{yagi10} 
Yagi, M. et al. 2010, AJ, 140, 1814
\bibitem[\protect\citeauthoryear{Yang et al.}{2007}]{yang07} 
Yang X., Mo H. J., van den Bosch F. C., Pasquali A., Li C., Barden M., 2007, ApJ, 671, 153
\bibitem[\protect\citeauthoryear{Yi et al.}{2005}]{yi05} 
Yi S.~K. et al., 2005, ApJ, 619, L111 
\bibitem[\protect\citeauthoryear{Zentner \& Bullock}{2003}]{zentner03} 
Zentner A. R., \& Bullock J. S., 2003, ApJ, 598, 49
\bibitem[\protect\citeauthoryear{Zentner et al.}{2005}]{zentner05} 
Zentner A. R., Berlind A. A., Bullock J. S., Kravtsov A. V., \& Wechsler R. H., 2005, ApJ, 624, 505 
\end{thebibliography}
\end{document}